# Bayesian Additive Regression Trees using Bayesian Model Averaging

Belinda Hernández[1,2], Adrian E. Raftery[1,3], Stephen R Pennington[2], and Andrew C. Parnell[1,4]

[1]*School of Mathematical Sciences, University College Dublin, Ireland*
[2]*School of Medicine and Medical Science, University College Dublin, Ireland*
[3]*Department of Statistics, University of Washington, USA*
[4]*Insight: The National Centre for Data Analytics, University College Dublin, Ireland,*
Email: belinda.hernandez@ucdconnect.ie

July 8, 2015

**Abstract**

Bayesian Additive Regression Trees (BART) is a statistical sum of trees model. It can be considered a Bayesian version of machine learning tree ensemble methods where the individual trees are the base learners. However for datasets where the number of variables $p$ is large (e.g. $p > 5,000$) the algorithm can become prohibitively expensive, computationally. Another method which is popular for high dimensional data is random forests, a machine learning algorithm which grows trees using a greedy search for the best split points. However, as it is not a statistical model, it cannot produce probabilistic estimates or predictions.

We propose an alternative algorithm for BART called BART-BMA, which uses Bayesian Model Averaging and a greedy search algorithm to produce a model which is much more efficient than BART for datasets with large $p$. BART-BMA incorporates elements of both BART and random forests to offer a model-based algorithm which can deal with high-dimensional data.

We have found that BART-BMA can be run in a reasonable time on a standard laptop for the "small $n$ large $p$" scenario which is common in many areas of bioinformatics. We showcase this method using simulated data and data from two real proteomic experiments; one to distinguish between patients with cardiovascular disease and controls and another to classify aggressive from non-aggressive prostate cancer. We compare our results to their main competitors. Open source code written in R and Rcpp to run BART-BMA can be found at: https://github.com/BelindaHernandez/BART-BMA.git

# 1 Introduction

Advances in technology and data collection have meant that many fields are now collecting and analysing bigger datasets than ever before (Lynch, 2008). This has brought the analysis of high-dimensional data to the forefront of statistical analysis (Bühlmann and Van De Geer 2011 ; Fujikoshi et al. 2011; Zhao et al. 2012). In many areas of research, especially biomedical applications, it is common to have datasets which collect very detailed data on a relatively small subset of observations, resulting in what is known as a "small $n$ large $p$" problem, where the number of variables $p$ is often much larger than the number of observations $n$. This precludes the use of many standard statistical techniques and means that only a restricted set of statistical models can be used (Hernández et al., 2014).

Random forests (RF), first proposed by Breiman (2001), is a popular method for dealing with high-dimensional data, mainly because of its computational speed and high accuracy. It is a non-parametric method and so does not make any major distributional assumptions about the data. RF automatically allows for non-linear interaction effects, a desirable property in many high-dimensional datasets (Nicodemus et al. 2010; Archer and Kimes 2008). The standard output of the RF method not only reports the accuracy of the algorithm, but also gives a variable importance measure for each variable which informs the user as to which variables were the most predictive. However, as RF is a machine learning algorithm and does not use a statistical model it cannot provide probability-based uncertainty intervals as in a Bayesian setting.

Bayesian methods have proven popular in many areas of research, in part because they are robust to over-fitting in the presence of small sample sizes and can handle missing or incomplete data. Because a full probability model is used, external information from, for example, previous experiments or academic literature can be incorporated into the model, which is an advantage over non-Bayesian statistical and machine learning techniques (Wilkinson, 2007). Bayesian methods also offer the flexibility to incorporate known experimental and biological variability into a prior probability distribution (Harris et al., 2009). A key benefit of using model-based approaches is that they give a principled estimate of uncertainty, which can be useful in decision-making. Machine learning algorithms usually provide point estimates only and so decisions are made ignoring the uncertainty surrounding these estimates.

Bayesian Additive Regression Trees (BART) (Chipman et al., 2010) is a Bayesian tree ensemble method similar in idea to gradient boosting (Friedman, 2001a), which combines the advantages of Bayesian models with those of ensemble methods such as RF. The advent of a parallelised R software package called `bartMachine` (Kapelner and Bleich, 2014a) has made BART a feasible option for analysis of a wide range of datasets. As BART is a model-based approach, it yields credible intervals for predicted values, in contrast to machine learning algorithms such as RF. However, as acknowledged in their manual (Kapelner and Bleich, 2014b), the algorithm for BART is very memory intensive for datasets with large numbers of variables. When tested on datasets with $p > 5,000$ `bartMachine` required a minimum of 10GB of RAM using its default settings.

In this article we propose an alternative algorithm for BART which we refer to as Bayesian Additive Regression Trees using Bayesian Model Averaging (BART-BMA). BART-BMA modifies the original BART method in a number of ways to make the algorithm more efficient for high-dimensional data. BART-BMA can be seen as a bridge between RF and BART in that it is model-based yet will run on high-dimensional data. One of the main reasons the original proposal of BART struggles in high dimensions is that it uses Markov Chain Monte Carlo (MCMC) to sample from the posterior distribution of the tree space. Rather than using MCMC and saving every iteration of the MCMC chain for each tree to memory, BART-BMA uses an efficient variant of Bayesian model averaging called Occam's window (Madigan and Raftery, 1994). This method discards models with low posterior probability and focuses final predictions on the subset of models with the highest posterior probabilities. In order to improve model selection speed, BART-BMA uses a greedy search algorithm to find predictive split points, so only high quality splits are proposed when growing tree models. Thus BART-BMA is computationally feasible for high-dimensional datasets, does not require any specialised hardware or software and brings with it the advantages of a model-based approach.

In this paper we showcase BART-BMA using a simulated example as well as two real applications to proteomic experiments. The article is organised as follows. Section 2 reviews existing tree based variable selection models such as RF (Breiman, 2001) and BART (Chipman et al., 2010). Section 3 describes our proposed model and explains the differences between it and BART. Section 4 compares BART-BMA to BART and their main competitor RF for a number of simulated datasets and applies these methods to two proteomics datasets. We conclude in Section 5.

## 2  Tree-Based Variable Selection

Tree-based models have long been used for prediction and classification, going back to 1963 for the analysis of survey data (Morgan and Sonquist 1963; Morgan 2005). Tree-based modelling came to the fore with the seminal work of Quinlan (1979) and Quinlan (1986), and particularly the Classification and Regression Tree (CART) method of Breiman et al. (1984).

Decision trees consist of internal nodes where splitting rules are imposed on the data in the form of $x_p \leq c$, where $x_p$ refers to the $p$th variable in dataset $X$ and $c$ is a threshold value within the range of values of variable $x_p$. Observations which satisfy the split rule are sent to the left hand daughter node and those which do not are sent to the right hand daughter node. Observations are iteratively split into left and right hand daughter nodes as they pass through each internal node in turn until a terminal node is reached.

One of the main reasons for the popularity of tree models over standard statistical models such as linear regression is that decision trees automatically search for and include non-linear interaction effects. It was later noted however that individual decision trees tend to over-fit and be sensitive to the training data they were built on. To counteract this, many ensemble methods were proposed where multiple diverse models are aggregated or averaged over to give a more stable and generalisable solution (Breiman 1996a; Breiman 1996b; Friedman 2001b).

### 2.1  Random Forests (RF)

Random forests (Breiman, 2001) is one of the most popular tree based ensemble algorithms and has been used in many fields (Ham et al. 2005; Svetnik et al. 2003; Daz-Uriarte and Alvarez de Andrs 2006). RFs use an average of multiple CART decision trees in their method. Each decision tree in the RF algorithm is based on a bootstrap sample of observations and trees are grown by splitting on a random sample of variables in each internal node. In this way RFs avoid over-fitting and give a cross-validated estimate of model performance by reporting accuracy on the out of bag samples that were not used to build the tree model.

Rather than using a statistical model, RF performs an exhaustive search for split points on various subsets of data. Each tree in the RF is grown to maximal depth so that each observation is predicted precisely in its terminal node. Therefore individual trees tend to be quite large and complex. The main reasons for the popularity of RFs are that they are generally very accurate, are computationally fast and will work for large datasets. Furthermore the RF algorithm provides a variable importance score for each variable used in the construction of the RF. The two main variable importance scores used are the decrease in Gini impurity which is generally used for classification problems and the mean decrease in accuracy which is generally used for regression problems. Therefore the RF can be used for model prediction and explanation unlike other black box algorithms such as support vector machines and neural networks which do not provide variable importance scores by default.

Another desirable aspect of RF is that it allows access to the trees that are averaged over in its model. RF however grows its trees to maximum depth and so tends to have very deep trees where many of the splits do not add to the predictive accuracy, thus making individual trees hard to interpret.

## 2.2 Bayesian Additive Regression Trees (BART)

BART is a Bayesian tree ensemble method where the response variable $Y$ is estimated by a sum of Bayesian CART trees (Chipman et al., 2010). Given an $n \times p$ matrix of explanatory variables $X$, let $x_k = [x_{k1}, \ldots, x_{kp}]$ be the $k$th row (i.e. the $k$th observation) of $X$. The basic BART model is

$$Y_k = \sum_{j=1}^{m} g(x_k; T_j, M_j) + \varepsilon_k, \tag{1}$$

where $g(x_k)$ is a Bayesian CART decision tree model as described in Chipman et al. (1998), $T_j$ refers to decision tree $j = 1 \ldots m$ where $m$ is the total number of trees in the model, $T_j$ has terminal node parameters $M_j$, and $\varepsilon \stackrel{\text{iid}}{\sim} N(0, \sigma^2)$ where $\sigma^2$ is the residual variance. The original BART model is fitted via a back-fitting Gibbs sampler which draws from the joint posterior distribution of all the trees and terminal node parameters and the standard deviation, given the data (Chipman et al., 2010).

Each tree $T_j$ is iteratively fit based on the residuals of the previous trees at the current iteration of the Gibbs sampler until a predetermined number of iterations is reached. Assuming prior independence of the trees $T_j$, the terminal node parameters $\mu_{ij} \in M_j$ (where $i = 1 \ldots b_j$ indexes the terminal nodes of tree $j$) and $\sigma$, the posterior distribution is

$$p(T, M, \sigma | X, Y) \propto p(Y|X, T, M, \sigma) \left[ \prod_j \prod_i p(\mu_{ij}|T_j) p(T_j) \right] p(\sigma), \quad (2)$$

where $p(Y|X, T, M, \sigma)$ is the overall likelihood of the sum of trees model. In (2), each tree $T_j$ has terminal node parameters $M_j$, $p(\mu_{ij}|T_j)$ is the prior distribution of the terminal node parameters $\mu_{ij} \in M_j$ given the tree structure $T_j$ and $p(\sigma)$ is the prior distribution of $\sigma$. $p(T_j)$ is the prior on tree $T_j$. BART places the same regularisation prior on the tree size and shape as that of Chipman et al. (1998) to deter any one tree from having undue influence over the sum of trees model. The probability that a given terminal node is further split into two children nodes is $P_{SPLIT} = \alpha(1 + d)^{-\beta}$, where $d$ is the depth of internal node $i$ and $\alpha$ and $\beta$ are parameters which determine the size and shape for tree respectively.

The prior distribution of the terminal node parameters $\mu_{ij}$ is $\mu_{ij}|T_j \overset{\text{iid}}{\sim} N(0, \sigma_0)$, where $\sigma_0 = \frac{0.5}{e\sqrt{m}}$ and $e$ is a user-specified parameter with recommended values between 1 and 3. To set the prior distribution of the $\mu_{ij}$, Chipman et al. (2010) noted that $\mathbb{E}(Y|X)$ is modelled as a sum of $m$ $\mu_{ij}$ parameters. BART then uses a data-informed prior for the $\mu_{ij}$ that implies that $\mathbb{E}(Y|X)$ lies within the range of values of $Y$ with high probability. Here the response variable $Y$ is centered at zero and scaled to have minimum and maximum values at $-0.5$ and $0.5$ respectively before analysis. For this reason the prior mean of $\mu_{ij}$ is set to 0. The prior terminal node variance $\sigma_0$ is then set such that $e\sqrt{m}\sigma_0 = 0.5$, and $e$ is chosen such that the prior implies that $\mathbb{E}(Y|X)$ lies between $-0.5$ and $0.5$ with high probability. For example $e = 2$ sets a 95% prior probability that the $\mathbb{E}(Y_{scaled}|X)$ lies between -0.5 and 0.5.

BART assumes that the model precision $\tau$ has prior distribution $\tau \sim Ga(\frac{\nu}{2}, \frac{\nu\lambda}{2})$, where $\tau = \sigma^{-2}$ with degrees of freedom $\nu$ and scale $\lambda$ respectively. This gives rise to the following full conditional distribution of the partial residuals $R_j|\ldots = Y - \sum_{l \neq j}^m g(X; T_l, M_l)$ for each tree $T_j$:

$$p(R_j|X, T_j, \tau) \propto \prod_{i=1}^b \left( n_i\tau + \frac{0.5}{e\sqrt{m}} \right)^{-\frac{1}{2}} \tau^{\frac{n+\nu}{2}-1}$$
$$\exp\left(-\frac{\tau}{2}\left(\sum_{\iota=1}^{n_i} R_{\iota ij}^2 + \nu\lambda\right)\right) \exp\left(\frac{n_i^2 \bar{R}_{ij}^2 \tau^2}{2\left(n_i\tau + \frac{0.5}{e\sqrt{m}}\right)}\right), \quad (3)$$

where $n_i$ is the number of observations in terminal node $i$ of tree $j$ and $\bar{R}_{ij}$ is the mean of the tree response variable $R_j$ for terminal node $i$. The termi-nal node parameters $M_j$ have been marginalised over in equation (3). This reduces the computational cost of BART as the calculation of the likelihood for each tree $T_j$ does not have to take account of the changing dimensionality of the terminal node parameters as trees are grown and pruned.

New trees in the BART algorithm are proposed in the Gibbs sampler by selecting from one of four proposal moves: GROW (node birth), PRUNE (node death), CHANGE (changing split rules) and SWAP (swapping internal nodes). The algorithm for BART is shown in Algorithm 1.

## 2.3 BART for Classification

For binary classification BART follows the latent variable probit approach of Albert and Chib (1993). Latent variables $Z_k$ are introduced so that

$$Y_k = \begin{cases} 1 & \text{if } Z_k > 0 \\ 0 & \text{otherwise.} \end{cases}$$

The sum of trees prior is then placed on the $Z_k$ so that $Z_k \sim N\left(\sum_{j=1}^{m} g(x_k; T_j, M_j), 1\right)$. It follows that that $Y_k$ is Bernoulli with

$$P(Y_k = 1|x_k) = \Phi\left(\sum_{j=1}^{m} g(x_k; T_j, M_j)\right), \qquad (4)$$

where $\Phi$ is the standard normal cumulative distribution function, used here as the link function. Note that there is no residual variance parameter $\sigma^2$ in the classification version of the model.

The full posterior distribution is then

$$p(T, M, Z|X, Y) \propto p(Y|Z)p(Z|X, T, M)\left[\prod_{j}\prod_{i} p(\mu_{ij}|T_j)p(T_j)\right], \qquad (5)$$

where the top level of the likelihood (i.e. the first term on the right hand side) is a deterministic function of the latent variables. The conditional prior distributions of the terminal node parameters $\mu_{ij}|T_j$ are set in the same way as that described in Section 2.2 except that $\sigma_0 = \frac{3}{e\sqrt{m}}$ instead of $\sigma_0 = \frac{0.5}{e\sqrt{m}}$. This is in order to assign high prior probability to the interval $(\Phi[-3], \Phi[3])$ which corresponds to the 0.1% and 99.9% quantiles of the normal c.d.f..

---
**Algorithm 1:** BART Algorithm
---
**Input**: $n \times p$ matrix $X$ with response variable $Y$
**Output**: Credible interval for $\hat{Y}$, after burn-in updates for $\sigma$
Initialize residuals $R_{00} = Y$
Set tree $T_0^0$ to a stump
**for** $\gamma \leftarrow 1$ **to** *niter* **do**
    **for** $j \leftarrow 1$ **to** $m$ **do**
        1. Generate a proposal tree $T^*$ by choosing from one of the following proposal moves:

            - GROW
            - PRUNE
            - CHANGE
            - SWAP

        2. Set $T_j^{\gamma+1} = T^*$ with probability
$\alpha\{T_j^\gamma, T^*\} = min\left\{\frac{q(T^*,T^\gamma)}{q(T^\gamma,T^*)} \frac{p(R_j|T^*,\sigma)p(T^*)}{p(R_j|T^\gamma,\sigma)p(T_j^\gamma)}, 1\right\}$
Else set $T_j^{\gamma+1} = T_j^\gamma$

        3. Update terminal node parameters $M_j$ by drawing from $p(M_j|T_j, R_j, \sigma)$

        4. Update $\sigma$ parameter by drawing from $p(\sigma|M, T, R)$

        5. Update predicted values for $T_j^\gamma$

        6. Update $R_{j+1} = Y - \sum_{l \neq j+1}^{m} g(X; T_l, M_l)$

    Set $\hat{Y}_\gamma = \sum_{j=1}^{m} g(X; T_j, M_j)$

**return** Credible interval for $\hat{Y}$, after burn-in updates for $\sigma$

---

The latent variables $Z_k$ introduce an extra step in the Gibbs algorithm to fit the model. The full conditional distributions of $Z_k|\ldots$ are:

$$Z_k|\ldots \sim \begin{cases} \min\left[N\left(\sum_j g(x_k; T_j, M_j), 1\right), 0\right] & \text{if } Y_k = 1, \\ \max\left[N\left(\sum_j g(x_k; T_j, M_j), 1\right), 0\right] & \text{if } Y_k = 0. \end{cases} \quad (6)$$

The other updates (e.g. for $T_j|\ldots$ and $M_j|\ldots$) are the same as in the standard BART algorithm of Section 2.2. The difference is that now the response is the latent variable $Z_k$, from which residuals $R_j$ can be calculated for updating of individual trees.

## 2.4 Issues with current tree-based methods

RF and BART are both popular and useful ensemble methods which not only provide a measure of accuracy but also a variable importance score which can be used for variable selection.

A major advantage of RF is that it has a fast running time and can be applied to high-dimensional datasets on a standard laptop computer. However, a disadvantage is that the method does not provide a principled assessment of uncertainty about the prediction. Only point predictions of $\hat{Y}$ are given with no estimate as to the variability of these predicted values.

BART on the other hand is a fully specified Bayesian model and so can provide estimates of model and predictive uncertainty. However, it becomes prohibitively memory intensive when used on high-dimensional data. Two main bottlenecks were noted in the BART model as described in Section 2 where $p$ is large. The first is that using a uniform prior to choose predictive split rules in each internal node of each tree produces chains with high rejection rates for large $p$. Thus the MCMC algorithm becomes inefficient, especially if the number of truly predictive variables is small. Another issue for high-dimensional data is that the BART algorithm is very memory hungry when the length of the chains and the size of each sum of trees are large (a by-product of $p$ being large). This is because each tree model for each iteration of each MCMC chain must be saved to memory if needed for prediction of an external dataset. All of these issues are addressed by BART-BMA.

# 3 BART-BMA

BART-BMA is a Bayesian CART ensemble model, which like BART uses a sum of trees model. BART-BMA can be seen as a bridge between RF and BART in that, like RF it can be applied to high-dimensional datasets on a standard laptop, but it is also a model-based sum of trees method like BART and therefore can provide estimates of predictive uncertainty.

BART-BMA departs from the original BART model in a number of ways. First BART-BMA performs a greedy search for predictive split rules and only grows sum of trees models based on this set of most predictive splits. Then, rather than using MCMC, BART-BMA uses an efficient implementation of Bayesian Model Averaging which averages over multiple models. Each model in BART-BMA consists of a sum of trees in a similar vein to BART. The algorithm for BART-BMA is shown in Algorithm 2.

## 3.1 Finding Good Split Rules

BART-BMA uses a greedy search method for finding predictive split points. The alternative of uniformly proposing split rules for tree internal nodes, used by BART, becomes very slow as the number of variables increases. It will be particularly inefficient when the number of variables associated with the response is low, as in many bioinformatics studies. With high-dimensional data we argue that a greedier search should be adopted in order to focus the algorithm towards predictive splitting rules. BART-BMA proposes the use of a change point algorithm called Pruned Exact Linear Time (PELT) as proposed in Killick et al. (2012) as well as a grid search method, both of which we now discuss.

### 3.1.1 PELT

Searching for predictive split points for a single variable in a tree is equivalent to searching for change points in a univariate stochastic process. Because of this, BART-BMA uses a change point detection algorithm called Pruned Exact Linear Time (PELT) to search greedily for predictive split points (Killick et al., 2012). PELT was originally proposed to detect change points in time series data by minimising the function

$$\min_{\delta} \left[ \sum_{i=1}^{m+1} \left[ C(y_{(\delta_{i-1}+1):\delta}) + D \right] \right], \tag{7}$$

where $C(\cdot)$ is a cost function (such as the negative log likelihood) of each segment $i$ containing observations $y_{(\delta_{i-1}+1):\delta}$, and $D$ is a penalty for adding additional change points. PELT extends upon the optimal partitioning method of Yao (1984) by eliminating any change points which cannot be optimal. This is achieved by observing that if there exists a candidate change point $s$ where $\delta < s < S$ which reduces the overall cost of the sequence then the change point at $\delta$ can never be optimal and so is removed from consideration (Killick et al., 2012).

BART-BMA implements PELT as an efficient algorithm to find good split rules for each tree in the model. This function detects changes in the mean or variance of the tree response variable $R_j$ with respect to each variable $x_p$ in turn, where each change point is treated as a potential splitting rule to grow a tree. The subset of the best `numcp`% split points and split variables are chosen on the basis of its residual squared error. Our experience of testing BART-BMA on a number of different datasets is that a value of $D = 10\log(n)$ performs well as a general default for the PELT penalty when the number of observations $n < 200$. For larger values of $n$ we recommend using a higher value for $D$ or the grid search option instead (see Section 3.1.2). BART-BMA implements a version of PELT in $C^{++}$ which is equivalent to the `PELT.meanvar.norm` function from the `changepoint` package in R (Killick et al., 2014). This function searches for changes in the mean and variance of variables which are assumed to be normally distributed. Additional change points are accepted if there is support for their inclusion according to the log likelihood ratio statistic.

### 3.1.2 Grid Search

An alternative search algorithm offered by BART-BMA is the grid search. Here each variable $x_p$ in dataset $X$ is split into `grid_size` + 1 equally spaced partitions within the range of $x_p$. Each partition value is then used as a potential split point and the best `numcp`% split rules in terms of the residual squared error are chosen. Increasing the `grid_size` parameter to a large number is not recommended where both $n$ and $p$ are large, as BART-BMA may become prohibitively slow. The value `grid_size` = 15 led to good performance in most of the cases we implemented.

### 3.1.3 Updating Split Rules

By default BART-BMA chooses the best `numcp`% of the total split rules before the tree is grown and only trees using the most predictive split rules are considered for inclusion. However the best split rules can also be updated for each internal node in the tree, similarly to how RF creates trees. We have found that updating split rules at each internal node generally results in smaller sum of trees models, but each tree within the sum tends to be larger and to choose splits that are similar to the primary splits of trees in the RF. We have found that updating the split rules at each internal node can in some cases increase the predictive accuracy, but generally at the expense of computational speed.

## 3.2 Terminal Node Prior

BART-BMA uses a different prior on the $\mu_{ij}$ parameters to that used in BART as described in Section 2.2. We use the standard conjugate prior suggested by Chipman et al. (1998) and assume $\mu_{ij}|T,\sigma \sim N(0, \frac{\sigma^2}{a})$. The reason BART-BMA chooses a different prior on $\mu_{ij}$ to BART is for computational simplicity as both the terminal node mean and the model precision $\tau = \sigma^{-2}$ can be marginalised out of the calculation of the tree likelihood, as shown in (8). This means that the model precision does not need to be updated through MCMC as in BART (see (3)).

## 3.3 BART-BMA Sum of Tree Likelihood

Unlike BART which calculates the likelihood for each individual Bayesian CART tree in the model, BART-BMA specifies the likelihood for each sum of trees in its model as follows:

$$p(Y|X,T) \propto \left[\nu\lambda - (Y^TW)(W^TW + aI)^{-1}(W^TY) + Y^TY\right]^{-\frac{n+\omega+\nu}{2}} \quad (8)$$

Here we define $J_j$ for each Bayesian CART tree $j = 1\ldots m$ in the sum of trees model as an $n \times b_j$ binary matrix where the elements of $J_j$ denote inclusion of observation $k = 1\ldots n$ in terminal node $i = 1\ldots b$ of tree $j$. We let $W = [J_1 \ldots J_m]$ be an $n \times \omega$ matrix, where $\omega = \sum_{j=1}^{m} b_j$, and $O = [M_1^T \ldots M_m^T]^T$ be a vector of size $\omega$ of terminal node means assigned to trees $T_1 \ldots T_m$. We then set $Y \sim N(WO, \tau^{-1})$ and $O \sim N(0, \frac{\tau^{-1}}{a}I)$. The parameters $\nu$, $\lambda$ and $\tau$ are specified as described in Section 2.2. As in Chipman et al. (2010) we shift and scale the response variable to have mean 0 and standard deviation 1 so $Y^TY = n$. When $m = 1$, (8) is equivalent to the full conditional distribution for an individual Bayesian CART model, as described by Chipman et al. (1998).

## 3.4 BART-BMA Method

BART-BMA grows Bayesian CART trees in the same manner as described in Chipman et al. (1998). The algorithm begins with a tree stump and only grows trees with high posterior probability.

For each tree in BART-BMA, split points are chosen for each variable in $X$ using one of the greedy search algorithms described in Section 3.1. BART-BMA starts by setting the `current_tree_list` to a stump tree which has all the data in one terminal node. It then iteratively grows the tree model $T_{hi\ell}$ for each terminal node $i$ in tree $\ell$ using each split rule $h$.

For each tree model in BART-BMA, the posterior probability of the sum of trees containing $T_{hi\ell}$, which we refer to as $ST_{hi\ell}$, is approximated using the BIC, defined as

$$BIC = -2(\log(p(Y|X, ST_{hi\ell})) + \log(p(ST_{hi\ell}))) + B\log(n). \qquad (9)$$

(Schwarz, 1978). In (9), $p(Y|X, ST_{hi\ell})$ is the tree likelihood for the sum of trees model containing $T_{hi\ell}$ as described in Section 3.3, $B$ is the number of parameters in the model, and $\log(p(ST_{hi\ell}))$ refers to the log tree prior for the sum of trees model. Here we use the same regularisation prior on the tree size and shape as Chipman et al. (2010) (see Section 2.2). Like BART, BART-BMA seeks to keep individual tree components small so that no one tree can dominate the overall model. Thus high prior probability is placed on trees with $< 5$ terminal nodes, as in Chipman et al. (2010). To calculate the prior probability of a given tree $T_{hi\ell}$, we set the prior $p(T_{hi\ell}) = \prod_{u=1}^{U} P_{SPLIT}$ for each internal node $u$ in the tree. The prior for $ST_{hi\ell}$ is then $\prod_{j=1}^{m} p(T_{hi\ell})$.

As the terminal node mean and variance parameters have been integrated out of (8), the only parameters remaining in the calculation of the tree likelihood are the split variable and the threshold at which that variable is split. As a split variable and split point are chosen at each internal node in the tree model, $B$ is equal to twice the number of internal nodes in the sum of trees model.

Trees are grown to a maximum user-specified depth and only splits that have at least 5 observations in the child nodes are retained, in order to avoid growing large trees with nearly empty nodes. Once a tree has been grown, the predicted values for observations falling in each terminal node $R_{hij}$ are set to the mean of the complete conditional of $p(\mu_{hij}|\ldots)$ as shown in Section 3.4.4.

### 3.4.1 Occam's Window

In order to focus the algorithm on highly probable tree models, Bayesian model averaging (BMA) is used. BMA takes account of model uncertainty by averaging over the posterior distribution of many models instead of focusing on the most probable one. Models are averaged over and weighted according to their posterior probabilities. In this way BMA takes account of the variability due to model selection.

As it is not possible to perform an exhaustive search of the model space especially when $p$ is large, we use a greedy and efficient version of BMA called Occam's Window (Madigan and Raftery, 1994). Here only the best subset of models which fall within Occam's Window are averaged over using (10).

$$\log(BIC_\ell) - \operatorname{argmin}_\ell(\log(BIC)) \leq \log(o). \tag{10}$$

As BART-BMA searches the model space, it approximates the posterior probability of each tree using the BIC (9). The lowest (best) BIC of any model encountered so far is saved and any sum of trees whose BIC falls within a given threshold $o$ of the best model is saved, while those outside of Occam's Window are discarded. Hence only those models for which there is high support from the data are maintained and those whose predictions are considerably worse than the best model are eliminated from consideration. Our experience has led us to use $o = 1,000$ as a general default value.

For each sum of trees model initially accepted in Occam's Window, the residuals are calculated and the process described in the previous paragraph is repeated using the partial residuals of each tree as the new tree response variable. This process of iteratively fitting trees is repeated until either a user-specified number of iterations is reached or no more trees are accepted in Occam's window. It should be noted that as models with better (lower) BIC are added to Occam's Window, this may eliminate some models which were previously within Occam's Window. Thus Occam's Window constantly updates the best list of sum of trees models as the algorithm proceeds. We have found that iteratively fitting trees with five trees in each sum of trees model generally performs well.

Once BART-BMA has selected the set of sums of trees within Occam's window, the predicted response values are then calculated as a weighted average of the predicted values from the selected sum of trees models. Each sum of trees model is weighted by its approximate posterior probability, $w_\ell / \sum_k w_k$, where $w_\ell$ is the model weight for model $\ell$, defined as

$$w_\ell = \exp\left(-0.5 \text{ BIC}_\ell - v\right), \tag{11}$$

where $v = \max_l(-0.5 \text{ BIC}_\ell)$. The BART-BMA algorithm is detailed in Algorithm 2.

### 3.4.2 Variable Importance

Like BART, BART-BMA provides a principled variable importance score, which is simply the estimated posterior expectation of the number of splitting rules involving the variable. For each sum of trees model $\ell$ with posterior probability $w_\ell$ as calculated in (11), let $\kappa_{p\ell}$ be the number of splitting rules

---
**Algorithm 2:** BART-BMA for continuous response
---
**Input**: $n \times p$ matrix $X$ with continuous response variable $Y$
**Output**: RMSE, Credible interval for $\hat{Y}$, after burn in updates for $\sigma$
Initialise: $Tree\_Response = Y\_scaled$; $num\_trees\_current\_round = 1$
Initialise: $current\_tree\_list = previous\_tree\_list =$ a tree stump
Initialise: $lowest\_BIC$

**for** $j \leftarrow 1$ **to** $m$ **do**

    **for** $\ell \leftarrow 1$ **to** $num\_trees\_current\_round$ **do**

        1. **Find Good Split Rules:**

        Run greedy search to find *numcp* best split rules for each tree $T_\ell$ in the *previous_tree_list*, save to matrix *splitting_rule*

        2. **Grow Greedy Trees in Occam's Window**

        **For** $H \leftarrow 1$ **to** $max\_tree\_depth$

        {

          Grow tree $T^*$ using each split rule in *splitting_rule* for each terminal node in each tree in *previous_tree_list*

          **If** Sum of trees including $T^*$ is in Occam's Window

            Append $T^*$ to *current_tree_list*

        }

        3. **Make sum of trees models and update residuals**

- Set *previous_tree_list*=*current_tree_list*
- Make list of sum of trees and save those in $Occam's\ Window$
- Update residuals for each sum of trees model and save to $Tree\_Response$ matrix
- $lowest\_BIC = min$(BIC of trees in *sum_tree_list*)

4. $\hat{Y} =$**Sum of weighted predictions over all sum of trees models**

5. **Implement post-hoc Gibbs Sampler for each sum of trees accepted in Occam's Window**

**return**:
    Credible intervals for $\hat{Y}$; Sum of trees in Occam's Window;
    Posterior probability of each sum of trees model
---

containing variable $x_p$ in model $\ell$. The variable importance score used by BART-BMA is then

$$\widehat{\mathrm{Imp}}(x_p) = \frac{\sum_{\ell=1}^{L} w_\ell \kappa_{p\ell}}{\sum_{p=1}^{P} \sum_{\ell=1}^{L} w_\ell \kappa_{p\ell}}. \tag{12}$$

### 3.4.3 Posthoc Gibbs Sampler

In order to provide credible intervals for the point predictions, $\hat{Y}$, provided by BART-BMA, a Gibbs sampler is run in a similar fashion to BART. For each sum of trees model accepted by BART-BMA a separate chain in the MCMC algorithm is run. Each terminal node parameter $\mu_{ij}$ in each tree $T_j$ is then updated followed by an update of $\sigma$. The details of the updates for the complete conditional of $p(\mu_{ij}|T_j, R_j, \sigma^2)$ and of $p(\sigma^2)$ are given in the following sections. The Gibbs sampler yields credible intervals for each set of sum of trees models accepted by BART-BMA along with the updates for $\tau = \sigma^{-2}$ for each set of trees accepted in the final BART-BMA model. The final simulated sample from the overall posterior distribution is obtained by selecting a number of iterations from the Gibbs sampler for each sum of trees model proportional to its posterior probability, and combining them. The post-hoc Gibbs sampler used by BART-BMA is far less computationally expensive than that of BART as it requires only an update for $\mu_{ij}$ and $\sigma$ from the complete conditional of each sum of trees model, which is merely a draw from a normal distribution and an inverse-Gamma distribution respectively (see Sections 3.4.4 and 3.4.5 respectively).

### 3.4.4 Update of $p(M_j|T_j, R_{\iota ji}, \sigma^2)$

Let $\mu_{ij} \in M_j$ index the $b$ terminal node parameters of tree $T_j$, and $R_{\iota ij}$ be the partial residuals for observations $\iota$ belonging to terminal node $i$ used as the response variable to grow tree $T_j$. BART-BMA assumes that the prior on terminal node parameters is $\mu_{ij}|T_j, \sigma \sim N(0, \frac{\sigma^2}{a})$, as in Chipman et al. (1998). The prior distribution of the partial residual is $R_j|\ldots \sim N(\mu_{ij}, \sigma^2)$.

The full conditional distribution of $M_j$ is then

$$\begin{aligned} p(M_j|T_j, R_{\iota ij}, \sigma) &\propto p(R_{\iota ij}|T_j, M_j, \sigma) p(M_j|T_j) \\ &\propto \prod_{\iota=1}^{n_i} p(R_{\iota ij}|T_j, M_j, \sigma) p(M_j|T_j), \end{aligned} \tag{13}$$

where $\iota$ indexes the observations within terminal node $i$ of tree $T_j$ and $n_i$ refers to the number of observations which fall in terminal node $i$.

The draw from the full conditional of $p(M_j|\ldots)$ is then a draw from the normal distribution

$$M_j|T_j, R_{\iota ij}, \sigma \sim N\left(\frac{\sum_{\iota=1}^{n_i} R_{\iota ij}}{n_i + a}, \frac{\sigma^2}{n_i + a}\right). \quad (14)$$

The full conditional of $M_j|\ldots$ depends only on $\sigma$ in the variance parameter, making it slightly more efficient than the update of $M_j$ using the BART prior which depends on $\sigma$ in both the mean and variance parameter.

### 3.4.5 Update of $p(\sigma^2)$

BART-BMA performs the update for $p(\sigma)$ in the same way as Chipman et al. (2010). The full conditional distribution of $\sigma^2$ is:

$$p(\sigma^2|R_j, T_j, M_j) \propto \prod_{k=1}^{n} p\left(R_j|T_j, M_j, \sigma^2\right) p\left(\sigma^2\right), \quad (15)$$

where $R_j \sim N\left(\sum_{j=1}^{m} g(x_k, T_j, M_j), \sigma^2\right)$ and $\frac{1}{\sigma^2} \sim \text{Gamma}(\zeta, \eta)$, where $\zeta$ and $\eta$ are equal to $\frac{\nu}{2}$ and $\frac{\nu\lambda}{2}$, respectively.

BART-BMA makes the draw for $\sigma^2$ in terms of the precision $\tau = \frac{1}{\sigma^2}$ where $p(\tau|R_j, T_j, M_j)$ is calculated as:

$$\tau|R_j, T_j, M_j \sim \text{Gamma}\left(\zeta + \frac{1}{2}, \frac{P}{2} + \frac{1}{\eta}\right), \quad (16)$$

where $P = \sum_k \left[Y_k - \sum_j g(x_k, T_j, M_j)\right]^2$. The next value of $\tau$ is then drawn from (16) and the value of $\sigma$ is calculated by getting the reciprocal square root of that value.

## 3.5 BART-BMA for Classification

BART-BMA can also be used for binary classification. We follow the same strategy as Section 2.3 by introducing the latent variables $Z_k \sim N(\sum_{j=1}^{m} g(x_k; T_j, M_j), 1)$ with

$$Y_k = \begin{cases} 1 & \text{if } Z_k > 0 \\ 0 & \text{otherwise,} \end{cases}$$

as before.

Our method requires estimates of $Z_k$ so that the previously introduced BART-BMA algorithm for continuous responses can be run without modification. We simply fix the $Z_k$ for all $k$ at the start of the algorithm. In practice we have found that this approach works well if we set all
$Z_k = \Phi^{-1}(0.001) \approx -3.1$ if $y_k = 0$ and $Z_k = \Phi^{-1}(0.999)$ if $y_k = 1$.

Once the $Z_k$ values are set, BART-BMA uses these as the new response for the tree, updates the residuals $R_j$ and iteratively fits trees as before. In order to set predicted $\hat{R}_j$ values in the terminal nodes, BART-BMA uses the mean of the full conditional for $M_j$, as given by (14). As in the continuous case this prior will shrink the terminal node means $\mu_{ij}$ towards zero thus ensuring that no one tree can dominate the model.

# 4 Results

In this section we compare BART-BMA to RF and BART for a number of simulated datasets and also to two real proteomic data sets for the diagnosis of cardiovascular disease and aggressive versus non-aggressive prostate cancer.

All results reported in this section were run on a HP Z420 Workstation with 32GB RAM. BART was run using the `bartMachine` R package (Kapelner and Bleich, 2014b) and RF was run using the `randomForest` R package (Liaw and Matthew, 2015). All methods and results were obtained using R version 3.2.0. Default values for all model parameters were used for each method.

## 4.1 Friedman Data

As in Chipman et al. (2010) we use simulated data based on Friedman (1991) to compare the results of BART-BMA, RF and BART. The original simulated dataset of Friedman (1991) had 5 uniform predictor variables $x_1 \ldots x_5$ where

$$y = 10 sin(\pi x_1 x_2) + 20(x_3 - 0.5)^2 + 10x_4 + 5x_5 + \varepsilon. \qquad (17)$$

In order to see how the three methods compared over various dataset sizes, five datasets were constructed by appending random noise variables to the truly important variables shown in (17) above, such that the total number of variables in each data set was $p = (100, 1000, 5000, 10000, 15000)$ where $x_1 \ldots x_p$ are uniform random variables and $\varepsilon \sim N(0,1)$. Hence each method was compared across 5 datasets using five-fold cross-validation and the cross-validated root-mean squared error (RMSE) was recorded along with the CPU time taken to run each method. For the Friedman datasets BART-BMA results are shown using the GRID search, as in all cases the number of observations is $n = 500$ (see Section 3.1.1 for default recommendations).

Figure 1 shows the results for the model accuracy (Figure 1a) and CPU time (Figure 1b) of the three methods for the simulated Friedman data. As can be seen, BART performed best with regards to RMSE for small datasets where $p < 1,000$ (Figure 1A). However, once the number of variables was increased past $p \geq 5,000$ BART-BMA outperformed both RF and BART. In fact BART-BMA outperformed RF and BART for all cases where $p > 1,000$ regardless of the method used for finding good splits or whether the split rules were updated before the tree or during the tree growing process.

With 5,000 variables, BART required a minimum of 10GB of RAM to run its default model, which is already beyond the capabilities of a current standard laptop. When $p = 10,000$ BART required a minimum of 22GB of RAM, and for $p = 15000$ it required a minimum of 30GB of RAM which would necessitate a specialised computer. Unlike BART, BART-BMA does not have any special memory requirements and will run on a standard laptop with 4GB of RAM for all five cases shown here.

With regard to model speed it can be seen from Figure 1B that for all 5 datasets BART-BMA was considerably faster than RF and was also more accurate for all but the case where $p = 100$. BART performed the fastest for all cases where $p > 1000$, but its model accuracy for these datasets was far worse than either BART-BMA or RF.

In order to assess how well calibrated the prediction intervals from BART-BMA and BART are, we show the average coverage of the in-sample 95% prediction intervals and the average interval width for each of the simulated datasets in Table 1. RF does not provide confidence intervals and so could not be included. If the method is calibrated, on average we would expect 95% of the intervals to contain the true value of $y$. From Table 1 it can be seen that BART-BMA is the best calibrated across the 5 simulated datasets shown, as its coverage is much closer to 95% than BART in all cases.

The right hand side of Table 1 shows the average interval width for both BART-BMA and BART. Here it can be seen that for small datasets where $p \leq 1,000$ BART had a shorter interval width, but this should be seen in the context of the fact that BART is also not well calibrated. For larger datasets where $p > 1000$ however, BART-BMA has a much shorter mean interval width than BART and had more accurate coverage. Overall BART-BMA was more accurate and better calibrated than BART, and needed much less computation than BART for data sets where $p > 1000$.

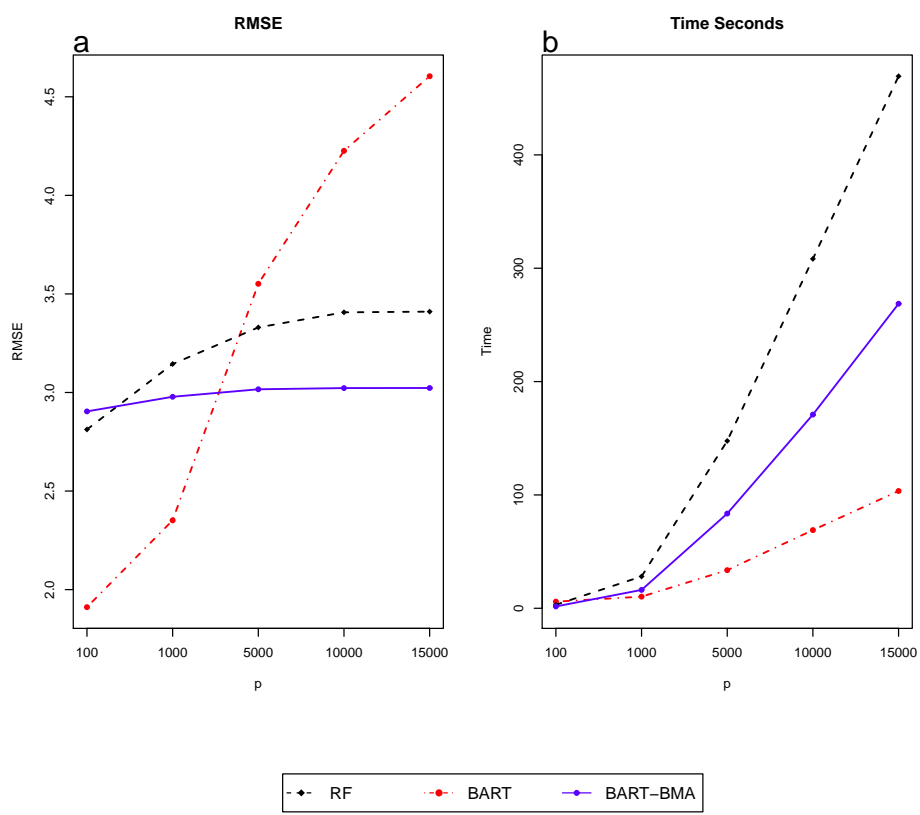

Figure 1: Friedman example: Comparison of RMSE and CPU run time in seconds for the 5 simulated Friedman datasets where $p = 100, 1000, 5000, 10000, 15000$. As $n = 500$ the GRID method was used to search for the subset of best splits. It should be noted although BART ran in the fastest time for $p \geq 5000$ it performed considerably worse than RF and BART-BMA for these datasets with respect to RMSE.

Table 1: Friedman example: coverage for in-sample 95% prediction intervals and average interval width for BART-BMA and BART (RF does not provide confidence intervals and so is not included). Perfect calibration is 95% hence the model with the lowest average interval width and a coverage as close to 95% as possible is most desirable. Items in **bold** refer to the best calibrated model with respect to interval coverage and interval width for each simulated dataset.

|       | Coverage | | Avg Int Width | |
|------:|:--------:|:----:|:-------------:|:----:|
| p     | BART BMA | BART | BART BMA | BART |
| 100   | **97.0%** | 99.7% | 11.73 | **7.19** |
| 1000  | **97.4%** | 99.9% | 11.69 | **9.90** |
| 5000  | **96.5%** | 100% | **11.67** | 15.39 |
| 10000 | **96.4%** | 100% | **11.66** | 17.46 |
| 15000 | **96.4%** | 100% | **11.68** | 18.07 |

### 4.1.1 Friedman Data: Variable Importance

We now analyse the variable importance scores for each method. As these data are simulated, it is known that variables $x_1 \ldots x_5$ are truly important and all other variables are random noise. As BART and BART-BMA provide variable inclusion probabilities, the decrease in Gini impurity provided by the RF was converted into a probability in order to allow for comparison between the variable importance scores assigned to each variable. The average importance for variables $x_1 \ldots x_5$ over the 5 cross-validation folds is reported in Figure 2.

BART-BMA had much larger average variable importance scores than RF and BART for variables $x_1$ and $x_2$ for all values of $p$. It also had much larger variable importance scores for $x_3$ when $p > 1,000$. For $x_4$, BART-BMA had substantial variable importance scores, as did RF. For $x_5$, BART-BMA had larger variable importance scores than the other methods for $p \leq 1,000$, and all three methods had low variable importance scores for $p > 1,000$. BART had strikingly low variable importance scores for the truly important variables, regardless of the numbers of noise variables.

Table 2 shows the sum of the variable importance scores assigned to the random variables $x_6 \ldots x_p$. Here it can be seen that across all five datasets BART-BMA correctly selected only those truly important variables in its model and never included any of the random noise variables. RF also con-siderably outperformed BART with regards to variable selection and assigned far lower importance to the random noise variables across all five datasets.

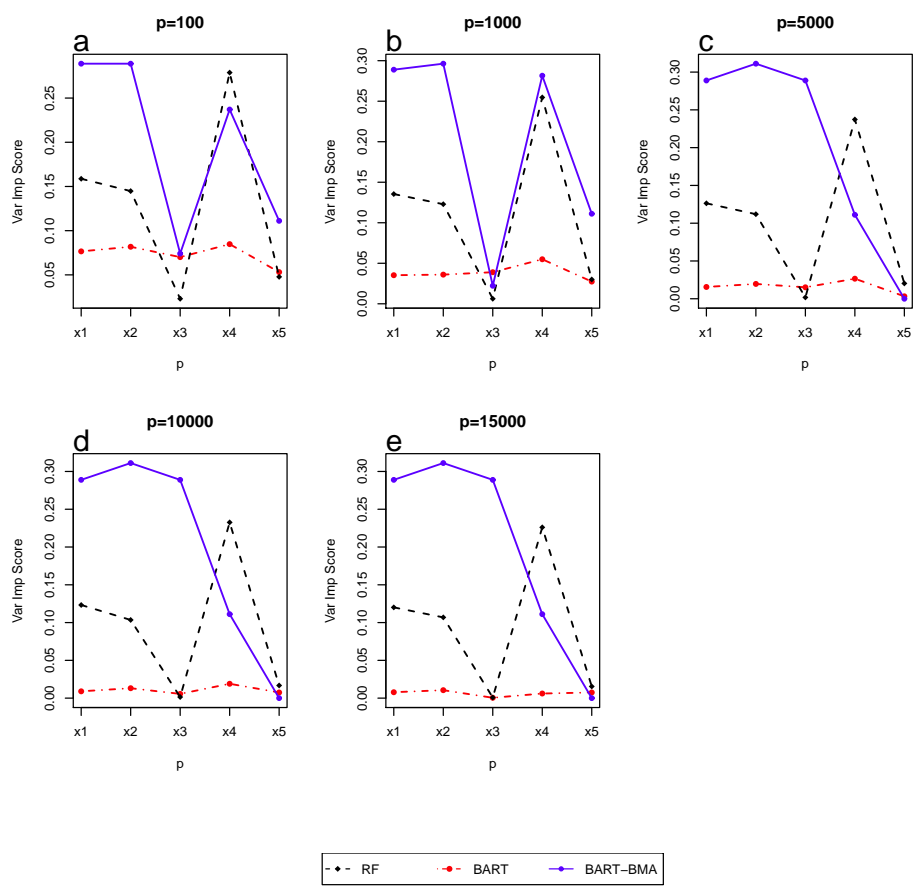

Figure 2: Friedman example: variable importance scores for the truly important variables $x_1 \ldots x_5$ for each of the 5 Friedman datasets. As $n = 500$ the GRID method was used to search for the subset of best split rules. BART-BMA and BART scores show the mean variable inclusion probability, RF scores show the mean decrease in Gini index expressed as a probability.

Table 2: Friedman example: Sum of the Variable Importance Score assigned to random variables $x_6 \ldots x_p$. The method which assigns the lowest importance to random variables is considered the best. Items in **bold** show the best method for each of the five simulated datasets.

| p | BART-BMA | RF | BART |
|---|---|---|---|
| 100 | **0** | 0.35 | 0.63 |
| 1000 | **0** | 0.45 | 0.81 |
| 5000 | **0** | 0.50 | 0.92 |
| 10000 | **0** | 0.52 | 0.95 |
| 15000 | **0** | 0.53 | 0.97 |

Table 3: Friedman example: Brier score $= \frac{1}{P} \sum_{p=1}^{P} (I_p - VIS_p)^2$ where $I_p = 1$ for truly important variables $x_1 \ldots x_5$ and $I_p = 0$ otherwise. Hence the lower the Brier score the better the model variable selection. Items in **bold** show the best model with respect to the Brier score.

| p | BART-BMA | RF | BART |
|---|---|---|---|
| 100 | **$3.24 \times 10^{-2}$** | $3.82 \times 10^{-2}$ | $4.30 \times 10^{-2}$ |
| 1000 | **$3.26 \times 10^{-3}$** | $4.00 \times 10^{-3}$ | $4.62 \times 10^{-3}$ |
| 5000 | **$6.55 \times 10^{-4}$** | $8.18 \times 10^{-4}$ | $9.68 \times 10^{-4}$ |
| 10000 | **$3.28 \times 10^{-4}$** | $4.13 \times 10^{-4}$ | $4.89 \times 10^{-4}$ |
| 15000 | **$2.18 \times 10^{-4}$** | $2.76 \times 10^{-4}$ | $3.29 \times 10^{-4}$ |

In order to further assess the overall quality of the variable importance scores assigned across all variables we used a Brier score $= \frac{1}{P} \sum_{p=1}^{P} (I_p - VIS_p)^2$ where $I_p = 1$ for truly important variables $x_1 \ldots x_5$ and $I_p = 0$ otherwise. Hence the model with the lowest Brier score is considered the best. Brier scores for all 5 simulated datasets are shown in Table 3. Here it can be seen that the variable importance scores from BART-BMA gave the best overall performance across all 5 datasets assessed and also that RF outperformed BART for each of the datasets with regards to the Brier score. Overall it can be seen that BART-BMA outperformed RF and BART with regards to the quality of its variable selection and did not include any random noise variables in its model regardless of the size of the dataset.

## 4.2 Prostate Cancer Data

Prostate cancer is a heterogeneous disease. In some men it manifests itself as an acute, aggressive and rapidly advancing condition, and in other men as a slow disease that is responsive to existing treatment regimes for significant periods of time. It is widely recognized that existing methods to classify the grade of the disease (using serum PSA levels, digital rectal examination and Gleason score) are not well suited for monitoring its progression nor establishing the optimal timing of treatment interventions (Logothetis et al., 2013). It is therefore important to be able to distinguish between aggressive and nonaggressive forms of the disease in a timely manner.

Here we show the results of an experiment where the expression levels of 51 peptides were measured using multiple reaction monitoring (MRM) on 63 patients with prostate cancer. Of these patients, 21 had extra capsular extension which is a surrogate for aggressive disease, while the remaining 42 had localised disease which had not spread beyond the boundaries of the prostate gland.

In order to access model performance we use the precision recall curve (Davis and Goadrich, 2006). The precision of a model is another term for the positive predictive value and measures the $\frac{\sum \text{True Positives}}{\sum \text{True Positives} + \sum \text{False Positives}}$. Here it is measuring the probability of a patient having extra capsular extension given that the model predicted they had extra capsular extension. The recall of a model is another name for the specificity and measures $\frac{\sum \text{True Negatives}}{\sum \text{True Negatives} + \sum \text{False Positives}}$. In this case the recall is measuring the probability of a patient being diagnosed as extra capsular extension by the model given they actually had extra capsular extension. The Precision Recall Curve shows the precision of a model over varying thresholds of the recall and the area under this curve is known as the Area under the Precision Recall Curve (AUPRC), the higher the AUPRC of a model the better the model.

Table 4 shows the comparison of BART-BMA, RF and BART for this dataset in terms of the classification rate, the AUPRC and the CPU time in seconds. BART-BMA using the PELT search performed best in terms of classification accuracy, identifying 79% of the cases correctly. It also performed best in terms of AUPRC with an area of 0.68. Here the PELT rather than the GRID search was used as the sample size for this experiment was small ($n \leq 200$).

Table 5 shows the five most important variables chosen across the three models. As can be seen here BART-BMA, RF and BART agreed that variables number 50 and 18 were the two most important for this dataset. RF and BART both chose variables 50, 18, 2 and 3 in their top five. BART-BMA tends to assign higher inclusion probabilities to a smaller number of variables than RF or BART which both tend to assign lower inclusion probabilities across a larger number of variables. In this case BART-BMA assigned a high variable importance score to variable 50 showing that this variable was present in 22.5% of the total splits across sum of trees models. If variables were chosen at random we would expect each variable to have an inclusion probability of 0.019 for this dataset and so across all cases a higher than random importance was assigned to the most important variables.

Table 4: Prostate cancer data: Classification rate, area under the precision recall curve (AUPRC) and CPU run time in seconds (standard deviation in brackets) for BART-BMA, RF and BART. Methods with higher classification rate, AUPRC and lower CPU running time are more desirable. Elements in **bold** show the best method with respect to each of the criteria.

|  | **BART-BMA** | **RF** | **BART** |
| --- | --- | --- | --- |
| Classification Rate | **0.79** | 0.71 | 0.71 |
| AUPRC | **0.68** | 0.67 | 0.65 |
| Time in Seconds | 1.02 (0.06) | **0.07 (0.01)** | 1.28 (0.61) |

## 4.3 Cardiovascular Disease

This section describes an experiment to distinguish patients with cardiovascular disease from control. This experiment was undertaken on 498 patients, 150 of whom had a cardiovascular disease and 348 of whom were healthy. A total of 36 proteins were measured by a targeted approach (MRM) in each patient sample. Table 6 shows the results for this experiment with respect to the classification rate, the AUPRC and the CPU time in seconds to run each method. BART-BMA was slightly more accurate than either RF or BART and correctly predicted 70% of the patients. With respect to the AUPRC however, BART-BMA did not perform as well as the other methods (see: Table 6) with BART performing the best by this measure. BART-BMA was quite fast in this instance and ran in 0.43 of a second, considerably faster than BART (4.82 seconds).

Table 7 shows the five most important variables chosen by each of the methods. As can be seen, all three methods chose quite similar proteins in this case with BART-BMA and RF agreeing on 4 of their top 5 and BART agreeing on three of its top 5. Again it can be seen that BART-BMA tends to assign a higher inclusion probability to a smaller set of proteins where as RF and BART tend to spread the probability across a larger number of variables. If the model were assigning probabilities uniformly across variables we would expect a probability $\sim 0.03$ which is the probability assigned to 4 of the top 5 variables included by BART and two of those included by RF. BART-BMA in all cases assigned a much higher than random inclusion probability to its most important variables.

Table 5: Prostate cancer data: Top five most important variables for each method. BART-BMA and BART scores show the mean variable inclusion probability, RF scores show the mean decrease in Gini index expressed as a probability. As $n < 200$ the PELT method was used to search for the subset of best split rules.

| Variable | BART-BMA | RF | BART |
|---|---|---|---|
| 50 | 0.225 | 0.082 | 0.025 |
| 18 | 0.168 | 0.052 | 0.024 |
| 2 |  | 0.035 | 0.021 |
| 3 |  | 0.042 | 0.022 |
| 4 |  |  | 0.021 |
| 30 | 0.103 |  |  |
| 31 | 0.071 |  |  |
| 25 | 0.050 |  |  |
| 44 |  | 0.037 |  |

Table 6: Cardiovascular disease data: Classification rate, area under the precision recall curve (AUPRC) and CPU run time in seconds (standard deviation in brackets) for BART-BMA, RF and BART. Methods with higher classification rate, AUPRC and lower CPU running time are more desirable. Elements in **bold** show the best method with respect to each of the criteria.

|  | BART-BMA | RF | BART |
|---|---|---|---|
| Classification Rate | **0.70** | 0.69 | 0.69 |
| AUPRC | 0.43 | 0.46 | **0.49** |
| Time in Seconds | **0.43 (0.22)** | 0.59 (0.04) | 4.82 (0.72) |

Table 7: Cardiovascular disease data: Top five most important variables for each method. BART-BMA and BART scores show the mean variable inclusion probability, RF scores show the mean decrease in Gini index expressed as a probability. As $n > 200$ the GRID method was used to search for the subset of best split rules.

| Variable | BART-BMA | RF   | BART |
|:--------:|:--------:|:----:|:----:|
| 14       | 0.33     | 0.05 | 0.04 |
| 4        | 0.16     | 0.03 |      |
| 24       | 0.14     | 0.05 | 0.03 |
| 3        | 0.06     | 0.03 | 0.03 |
| 34       | 0.06     |      |      |
| 2        |          |      | 0.03 |
| 10       |          |      | 0.03 |
| 13       |          | 0.03 |      |

# 5 Conclusions

We have proposed a Bayesian tree ensemble method called BART-BMA which modifies the BART method of Chipman et al. (2010) and can be used on datasets where the number of variables is very large. Instead of estimating the tree node parameters using MCMC, BART-BMA uses a version of Bayesian model averaging which is more memory efficient, and generally more accurate when the number of variables is large, as only a subset of the best models are averaged over. Changing the model priors to those used in Chipman et al. (1998) means that the model precision is needed only for calculation of the prediction credible intervals and not in the calculation of the likelihood. As such, a fast posthoc Gibbs sampler can be run, yielding estimates of predictive uncertainty in addition to point predictions.

BART-BMA proposes an efficient strategy for finding good split rules which works particularly well for high-dimensional data where the uniform prior used by BART becomes prohibitively computationally intensive. BART-BMA borrows elements of both BART and RF in that it is a sum of trees ensemble model which averages over multiple sums of trees and as such offers a model-based alternative to machine learning methods for high-dimensional data where BART is not feasible.

BART-BMA like RF not only provides a variable importance score but also provides access to the trees chosen in the final model. In general BART-BMA tends to choose shallower and more interpretable trees than RF as only splits which result in a high posterior probability are included.

We have showcased BART-BMA using both simulated and real life proteomic datasets and have shown its usefulness for high-dimensional data where BART will require a specialised computer or not run at all. We have found that BART-BMA is competitive with its main competitors RF and BART in terms of speed and accuracy and will run on a current standard laptop (4–8GB RAM).

We envisage future applications and extensions for BART-BMA including dealing with missing data as well as for use on longitudinal data and extending this model to the family of generalised linear models. BART-BMA could also be parallelised to reap further gains in computational speed.

**Acknowledgements:** We would like to thank Drs Chris Watson, John Baugh, Mark Ledwidge and Professor Kenneth McDonald for kindly agreeing for us to use the cardiovascular dataset described. Hernández's research was supported by the Irish Research Council. Raftery's research was supported by NIH grants nos. R01-HD054511, R01-HD070936, and U54-HL127624, and by a Science Foundation Ireland E.T.S. Walton visitor award, grant reference 11/W.1/I2079. Protein biomarker discovery work in the Pennington Biomedical Proteomics Group is supported by grants from Science Foundation Ireland (for mass spectrometry instrumentation), the Irish Cancer Society (PCI11WAT), St Lukes Institute for Cancer Research, the Health Research Board (HRA_POR/2011/125), Movember GAP1 and the EU FP7 (MIAMI). The UCD Conway Institute is supported by the Program for Research in Third Level Institutions as administered by the Higher Education Authority of Ireland.